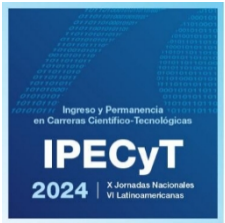
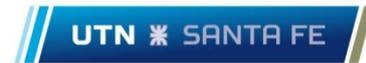

# Algoritmos de Resolución de Problemas como disparadores de procesos de carácter compensatorio y de reconstrucción de saberes previos

## Problem-Solving Algorithms as triggers of compensatory processes and reconstruction of previous knowledge

Presentación: 10/06/2024

Rodrigo E. Menchón

Facultad de Ciencias Exactas, Ingeniería y Agrimensura (FCEIA), Universidad Nacional de Rosario (UNR), Rosario, Argentina
Instituto de Física Rosario (IFIR), Rosario, Argentina
E-mail: menchon@fceia.unr.edu.ar

Santiago H. Luna

Secretaría de Investigación, Universidad Nacional de Hurlingham (UNaHur), Hurlingham, Argentina.
E-mail: santiago.luna@unahur.edu.ar

Andrea Fourty

Facultad de Ciencias Exactas, Ingeniería y Agrimensura (FCEIA), Universidad Nacional de Rosario (UNR), Rosario, Argentina
E-mail: fourty@fceia.unr.edu.ar

Hugo D. Navone

Facultad de Ciencias Exactas, Ingeniería y Agrimensura (FCEIA), Universidad Nacional de Rosario (UNR), Rosario, Argentina
Instituto de Física Rosario (IFIR), Rosario, Argentina
E-mail: hnavone@fceia.unr.edu.ar

**Resumen**

En este trabajo se presenta una estrategia de enseñanza implementada en Introducción a la Física, correspondiente al primer año del Profesorado en Física de la Universidad Nacional de Rosario, cuyo propósito principal es brindar a los estudiantes herramientas para comprender enunciados de problemas y ejercicios y para incorporar hábitos que favorezcan su resolución y su comunicación. Para esto, se implementó el uso de ciertos algoritmos de resolución de problemas, a los que denominamos Algoritmos-Rayuela, apelando a la imagen de este juego popular en el que los pasos se pueden saltar, alterar en su orden o ejecutar simultáneamente. Se pretende estimular un método de trabajo de manera crítica y problematizada, evitando que la aplicación de pasos de resolución se interprete rígida y dogmáticamente. Los testimonios recabados indican que la estrategia fue valorada positivamente por lxs estudiantes.

**Palabras clave:** Física Educativa, Algoritmos de Resolución de Problemas, Formación Inicial Docente, Lectoescritura Científica.

**Abstract**

In this work, we present a teaching strategy implemented in Introduction to Physics, corresponding to the first year of the Physics Teacher Degree at the National University of Rosario, whose main purpose is to provide students with tools to understand problem statements and exercises and to incorporate habits that favor their resolution and communication. For this purpose, we implemented the use of certain problem-solving algorithms, which we call Hopscotch-Algorithms, appealing to the image of this popular game in which steps can be skipped, rearranged or simultaneously executed. The aim is to stimulate a work method in a critical and problematized way, avoiding rigid or dogmatic applications of resolution steps. The testimonies collected indicate that the strategy was positively valued by the students.

## Introducción

Docentes, investigadores y autoridades educativas coinciden en identificar al primer año de la universidad como un periodo crucial, en el que la permanencia de gran parte de lxs estudiantes se ve amenazada. La transición de la escuela secundaria a la universidad es un proceso complejo en donde la autoconfianza se tensiona, los saberes previos resultan ser insuficientes, y se abordan nuevas formas de relacionarse con el conocimiento, lxs compañerxs y docentes.

Debido a que existe una gran heterogeneidad en las diferentes trayectorias educativas con las que cuentan lxs distintxs estudiantes de un mismo curso de primer año de la universidad, es que resultan necesarios dispositivos que promuevan procesos de carácter compensatorio y de reconstrucción de saberes previos, que contribuyan a fortalecer la autoconfianza en la propia singularidad y que faciliten el desarrollo de la capacidad de regulación autónoma del aprendizaje en cada unx de lxs estudiantes (Gimeno Sacristán y Pérez Gómez, 2002).

En este contexto, hemos incorporado Algoritmos de Resolución de Problemas en las clases de Introducción a la Física, unidad curricular del primer cuatrimestre del primer año del Profesorado en Física. En esta unidad curricular se abordan simultáneamente contenidos de Naturaleza de la Ciencia como una introducción a conceptos de Mecánica Clásica vectorial.

En el presente trabajo, tomamos al término "algoritmo" en su sentido amplio como un conjunto ordenado y finito de operaciones que permite hallar la solución de un ejercicio o problema. Los Algoritmos de Resolución de Problemas se incluyeron y presentaron junto con el resto de los contenidos del espacio curricular.

En la bibliografía se ubica una amplia variedad de diversos algoritmos de resolución de problemas propuestos por distintos autores. Quizás uno de los primeros a considerar sea Pólya (1945), y más cercanos en el tiempo y dentro de la Física se ubican Morin (2008) (véase p. 1), Sears, Zemansky y colaboradores (2009) (véase p. 2 y 3, o para Cinemática, Estática y Dinámica p. 7, 51, 82, 92, 137, 143), entre otros.

Entel (1988) señala que "(…) en la cultura escolar parecen predominar los conceptos de conocimiento como entidad abstracta, como instrumento para llegar a una verdad, como conjunto de contenidos organizados en una estructura". Siguiendo a Brovelli (2003) es posible señalar una correlación entre las concepciones de conocimiento que lxs docentes poseen y las estrategias de enseñanza y aprendizaje que llevan a cabo. En disciplinas como las Ciencias Empírico-Analíticas, la concepción hegemónica de conocimiento es considerarlo como entidad, como algo dado. Como alternativa superadora, se encuentra la concepción del conocimiento como producto de un proceso. En este posicionamiento:

> El conocimiento se basa en la percepción inicial del todo y concibe a cada elemento y a la totalidad como producto de un proceso. Atiende a las contradicciones considerándolas como motor de cambio. La actividad del sujeto al conocer supone implicación en el aprendizaje, lo que le permite reconocerse como partícipe de las transformaciones, construyendo o reconstruyendo el conocimiento. Implica estrategias de enseñanza y de aprendizaje de carácter reflexivo y crítico. (Brovelli, 2003)

Por otra parte, algunas acciones-herramientas, competencias y habilidades con las que deseamos que lxs estudiantes cuenten al finalizar su trayecto por el espacio curricular Introducción a la Física son, sin pretensiones de exhaustividad:

• interpretación y análisis de enunciados;

• identificación de ideas claves en textos;

• manejo de conceptos clave en Física como: sistema de coordenadas, instante inicial de referencia, leyes horarias y funciones de movimiento, diagramas de interacciones;

• comprensión de la coherencia subyacente entre: sistema de referencia elegido, esquema, funciones de movimiento y gráficas de las funciones de movimiento;

• desarrollo algebraico simbólico en la resolución de ejercicios;

• orden y claridad a la hora de elaborar un texto comunicando el desarrollo de una resolución de ejercicios de Física;

• disfrutar del proceso de estudiar Física;

• concebir a la Física como parte de la Cultura y las Ciencias.

Por otra parte, consideramos importante señalar que varias de estas acciones-herramientas del listado coinciden con la concepción de hablar ciencia de Lemke:

> El aprendizaje de la ciencia implica *aprender a hablar* en el idioma propio de ésta. Implica también utilizar este lenguaje conceptual tan especial al leer y escribir, al razonar y resolver problemas y durante la práctica en el laboratorio y en la vida cotidiana. Implica aprender a comunicarse en este idioma y ser miembro activo de la comunidad de personas que lo utilizan. 'Hablar científicamente' significa observar, describir, comparar, clasificar, analizar, discutir, hipotetizar, teorizar, cuestionar, retar, argumentar, diseñar experimentos, llevar a cabo procedimientos, juzgar, evaluar, decidir, concluir, generalizar, divulgar, escribir, disertar, y enseñar en y mediante el idioma de la ciencia. ¿Como aprendemos a hablar este idioma? aprendemos más o menos de la misma manera en que aprendemos cualquier otro idioma: practicándolo con las personas que lo dominan y empleándolo en las muy diversas situaciones donde se utiliza. (Lemke, 1997, p. 17)

En el presente trabajo, proponemos el estudio y uso de algoritmos para la resolución de ejercicios de Física como un dispositivo didáctico que permite incorporar al trabajo en la asignatura la adquisición de las habilidades anteriormente mencionadas.

Al momento de presentar los algoritmos de resolución, se pretende evitar una visión prescriptiva, rígida, acrítica de los mismos. Es por ello que se presenta como recomendación el uso de los mismos, considerando el orden de los pasos de resolución con flexibilidad. Para incorporar esta concepción al dispositivo, hemos llamado a los mismos Algoritmos-Rayuela. Esta imagen hace referencia a la posibilidad de que, a veces, dos o más pasos se den simultáneamente o se altere el orden de los mismos. De este modo, se pretende transmitir una visión compleja, problematizada, de los algoritmos de resolución: a la vez que se listan pasos de resolución, herramientas y procedimientos, se relativiza la rigidez con la que éstos pueden ser interpretados, evitando así una lectura dogmática de los mismos.

## Desarrollo

Dado que el propósito de esta propuesta es que los Algoritmos-Rayuela contribuyan a que los estudiantes adquieran principalmente la habilidad de organizar su escritura y saberes a la hora de enfrentar la resolución de ejercicios; pero a la vez se espera que no se constituyan en una receta con pasos a seguir acríticamente, se los presenta como sólo uno de los muchos posibles que ayudan a abordar la resolución de problemas. El primer Algoritmo-Rayuela de Resolución de ejercicios se presentó durante la primera clase. Se comentó y explicó la intensión y características de cada uno de sus pasos, a la vez que se contestaron dudas surgidas al respecto. A continuación, se presentan y comentan los pasos del correspondiente a la primera unidad de la asignatura.

Los pasos del Algoritmo-Rayuela de Resolución de ejercicios de cinemática son:

• 1) Leer detenidamente todo el enunciado del problema. Leer y releer el enunciado, prestando atención a cuáles son los datos y cuáles las incógnitas u objetivos.

• 2) Realizar el esquema inicial del problema. Identificar el tipo de problema, dibujar o esbozar la trayectoria del móvil, relevar datos e información conocida.

• 3) Definir el sistema de referencia a usar en el problema e incorporar su esquema. Haciendo foco en una concepción espaciotemporal del sistema de referencia: identificar instante inicial de referencia y origen del sistema de coordenadas. Identificar ejes del sistema de coordenadas x, y, z y sentido positivo de cada eje. Identificar posiciones relevantes e instantes relevantes (que son incógnitas o datos).

• 4) Considerar qué leyes horarias entran en juego en este problema. Escribir la forma general de las leyes horarias y las ecuaciones de velocidad y aceleración para el tipo de movimiento, esto es, las expresiones completas con todos los términos y variables.

• 5) Identificar herramientas que conecten los datos iniciales con el proceso de resolución que lleve al objetivo del problema. Aquí reescribir las leyes horarias y ecuaciones de velocidad y aceleración ahora en su forma particular para este problema, en función de los datos del enunciado, reemplazando cada variable por una expresión función de los símbolos presentes en el enunciado.

• 6) Desarrollar el proceso de resolución. Considerar desarrollos algebraicos, o procedimientos que permitan deducir la expresión objetivo del ejercicio, partiendo de las funciones de movimiento escritas en su forma particular.

• 7) Realizar las gráficas en tándem para las funciones de movimiento. Crear las gráficas de posición en función del tiempo, velocidad en función del tiempo y aceleración en función del tiempo, una sobre la otra, con el eje de las abscisas en común. Esto facilita una comparación rápida a la vez que refuerza la conexión existente entre las tres funciones de movimiento.

• 8) Corroborar la solución y verificar que no haya errores. Chequear expresiones finales, verificar unidades, corroborar pasos algebraicos. Responder ¿Tiene sentido el resultado? ¿Está completa mi respuesta?

• 9) "Agotar la exploración el problema". Considerar preguntas alternativas, pensar situaciones límite en expresiones finales, imaginar ejercicios análogos o similares.

• 10) ¡Disfrutar! Sentir alegría y placer de un ejercicio finalizado y explorado. Y compartir el proceso de resolución al colaborar con compañerxs de estudio, preguntando y/o explicando.

Para el paso 7) de gráficas de ejercicios de cinemática, coincidimos con Cabrera (2010 y 2012), cuando sostiene:

> Cuando los gráficos se construyen en tándem en un sólo golpe de vista obtenés la información de las tres magnitudes importantes en la cinemática: posición, velocidad y aceleración. El golpe de vista se realiza de esta manera: con una recta imaginaria vertical que atraviesa los tres gráficos. Por eso es importante que los 3 gráficos tengan la misma escala de tiempo. El orden de los gráficos también es importante. No sólo refleja la importancia de cada magnitud sino su orden de derivada: la velocidad es la derivada de la posición, y la aceleración es la derivada de la velocidad.

Hemos considerado importante incluir el último paso 10) ¡Disfrutar! debido a que creemos que tanto el interés, como el disfrute y una relación vital con el conocimiento son actitudes que se profundizan y cultivan (o no) a cada momento. La inclusión de este paso no debe ser interpretada como un impertativo hiperhedonista, sino más bien como un movimiento de apertura al deseo de saber, la formación, la educación, la humanización de la vida; un acercamiento hacia el conocimiento y hacia la cultura (Recalcati, 2016). En el juego tradicional de la rayuela, el objetivo es alcanzar el cielo, que es el último casillero. En cierto modo, uno de los objetivos de los Algoritmos-Rayuela estriba en comprender de qué modo cultivar el disfrute a través de los procesos de resolución de ejercicios y el estudio.

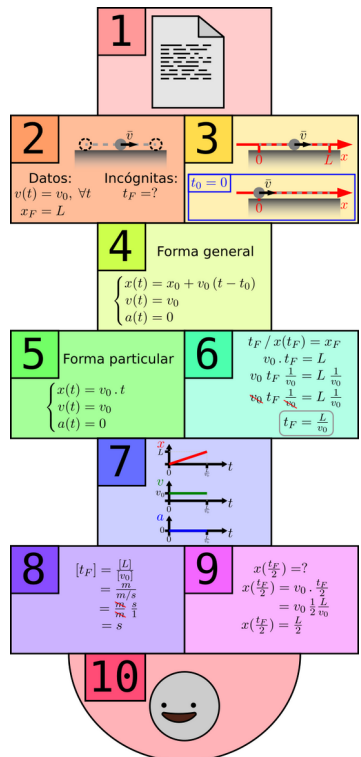

*Figura 1:* Representación esquemática de los pasos del Algoritmo-Rayuela de cinemática.

Luego de la presentación, se ilustró un proceso de resolución referenciando los pasos del Algoritmo-Rayuela en cada instancia. Desde la cátedra se ha relevado que el impulso inicial en los procesos de resolución de ejercicios de Física es el reemplazo de variables por sus valores numéricos. En este aspecto, coincidimos con Cabrera (2012) cuando dice que en las escuelas secundarias argentinas "Cuando un estudiante se topa con una letra a la que puede reemplazar por un número (y su unidad, si se tratase de una magnitud física) lo hace inexorablemente."

Considerando que uno de nuestros objetivos es fomentar el desarrollo algebraico simbólico, el primer ejercicio empleado para ilustrar la implementación de los diferentes pasos del Algoritmo-Rayuela se resolvió dos veces: la primera vez reemplazando cada variable por su valor numérico (imitando el modo de trabajo que en general resulta más conocido por lxs estudiantes) y la segunda vez se reinició todo el planteo pero en cada instancia se operó algebraicamente con las expresiones simbólicas (introduciendo el modo de trabajo hacia el cual se pretende llegar). Posteriormente, el equipo docente y lxs estudiantes sostuvieron un diálogo donde se analizaron comparativamente ambos modos y se enunciaron las ventajas del manejo algebraico simbólico.

Coincidimos con Morin (2008) cuando sostiene que el método de resolución simbólica algebraica presenta las siguientes ventajas respecto de la resolución numérica:

• es más rápido,

• es menos proclive a la aparición de errores,

• se resuelven infinitos problemas a la vez (para cada uno de los infinitos conjuntos de parámetros),

• facilita ver la dependencia de las incógnitas en función de las variables dato,

• se pueden estudiar casos límite,

• permite corroborar unidades

En cada ejercicio resuelto en el pizarrón durante las clases, se fueron señalando los pasos de resolución, siempre referenciándolos con los del Algoritmo-Rayuela, junto con comentarios, intercambio de ideas, discusiones. De este modo, se facilita el proceso de abstracción, a la vez que se crea el hábito de abordar los problemas de manera ordenada y reflexiva. Simultáneamente, se transmite una visión flexible y antidogmática sobre los algoritmos de resolución al visitar distintas implementaciones, alterando el orden de los pasos o adaptando los pasos a cada ejercicio.

Para cada tema se reelaboró, enriqueció y modificó el Algoritmo-Rayuela, con el fin de adaptarlo al tipo de ejercicio de cada campo de la Mecánica Clásica vectorial: Cinemática, Estática y Dinámica.

Siendo que en "Introducción a la Física" se abordan simultáneamente contenidos de Naturaleza de la Ciencia y de Mecánica Clásica vectorial, resulta pertinente el trabajo en torno a figuras y viñetas históricas. En palabras de Lemke:

> Es peligroso para la sociedad tener alumnos que dejan la escuela creyendo que la ciencia es una vía perfecta hacia lo absoluto, hacia las verdades objetivas, descubierta por gente con inteligencia sobrehumana. Aparte del peligro de que los hallazgos científicos sean utilizados para justificar políticas sociales perversas, una visión inhumana e impersonal puede alejar a muchos alumnos de la materia. Si deseamos animar a toda clase de alumnos a que se interesen en la ciencia y la usen para sus propios fines, necesitamos enseñarles lo que realmente es.
>
> *Los profesores han de enfatizar el lado humano de la ciencia: actividad real hecha por seres humanos reales, tanto en la época contemporánea como en períodos específicos de la historia. Las características personales de los científicos, con las que los alumnos puedan identificarse, deben ser puestas de relieve en vez de presentar a los científicos como superhumbres o extraterrestres*. (Lemke, 1997, p. 187-188)

Así, una actividad de trabajo alrededor del concepto de algoritmos es la lectura de un texto, junto con la visualización de un video (BBC ideas, 2019) y posterior debate acerca de los orígenes de los términos algoritmo y álgebra, rescatando la obra de Muḥammad Al-Jwarizmī. De este modo, el foco de las viñetas históricas pierde eurocentrismo y pone en valor los aportes del califato abasí, las culturas árabes y persas y la edad de oro del Islam.

Durante el primer parcial, no sólo se permitió sino que se impulsó a lxs estudiantes a que llevaran impreso el Algoritmo-Rayuela y que lo usaran durante el mismo.

Al iniciar la clase siguiente al examen, se repartió un cuestionario entre lxs estudiantes donde, entre otras, se les formuló la pregunta "En tus palabras, ¿para qué dirías que sirve el Algoritmo-Rayuela?". A continuación algunos resultados obtenidos, en palabras de nuestrxs estudiantes:

Estudiante A: "El Algoritmo-Rayuela es de mucha ayuda para tener una serie de pasos a seguir al resolver los problemas, de manera que la información y la resolución queden de forma prolija, ordenada y fácil de leer y de entender."

Estudiante B: "El Algoritmo-Rayuela sirve para ordenar la información y acordarse de qué paso sigue para resolver el ejercicio."

Estudiante C: "El Algoritmo-Rayuela sirve para ordenar la información en la hoja y que los contenidos sean más fáciles de entender para quien luego lo lea y también en el momento de resolver el ejercicio."

Estudiante D: "El Algoritmo-Rayuela es una manera de resolver ejercicios en general, sirve para muchas cosas."

Estudiante E: "El Algoritmo-Rayuela es un sistema ordenado y puesto en palabras, de un método para resolver ejercicios, específicamente de Física, pero se puede reformular de cierta forma para darle uso en otro tema o en otras situaciones."

Estudiante F: "El Algoritmo-Rayuela me ayuda a ubicarme mejor. Si me siento perdida o siento que algo me faltó, recurro al algoritmo muchas veces. Después de tanto usarlo ya no me hacía tanta falta porque me lo iba acordando."

Estudiante G: "El Algoritmo-Rayuela sirve para ordenarte un poco con los pasos que tenés que hacer y hace que sea más llevadera la resolución de un ejercicio."

Estudiante H: "Para mí, el Algoritmo-Rayuela sirve para ser más ordenado a la hora de resolver ejercicios, además de no olvidar ningún punto importante para completar el mismo."

Estudiante I: "El Algoritmo-Rayuela sirve para una mejor organización al momento de realizar un ejercicio y para que alguien ajeno al tema pueda comprenderlo."

Por otro lado, hemos notado que las resoluciones de los ejercicios en el examen parcial se corresponden con lo trabajado en clases, observándose que éstos se desarrollan, en general, de manera ordenada, con los gráficos correspondientes, mostrando que los estudiantes van adquiriendo paulatinamente los hábitos que deseamos fomentar. Asimismo, se considera que la implementación del Algoritmo-Rayuela contribuye a la mejora de las habilidades comunicacionales de lxs estudiantes. Se observa además que lxs estudiantes han logrado cultivar sus competencias de desarrollo algebraico simbólico y que sólo efectúan los cálculos numéricos en instancias finales de sus planteos. Esto también contribuye al orden y claridad de la escritura de la resolución de los ejercicios.

## Conclusiones

En este trabajo se ha presentado una estrategia didáctica dirigida a estudiantes de primer año del Profesorado en Física buscando favorecer la adquisición de habilidades metodológicas y de comunicación en la resolución de ejercicios y problemas.

Para ello, se ha propuesto la utilización sostenida de Algoritmos de Resolución de problemas. Para evitar caer en la aplicación de los mismos como mera receta, los hemos llamado "Algoritmos-Rayuela" enfatizando, con su nombre y con su implementación, las características críticas, reflexivas, dinámicas y antidogmáticas de los mismos. Estos algoritmos presentan características propias según el tema a trabajar.

Durante las clases, la resolución de los ejercicios se llevó a cabo haciendo permanentemente referencia a los pasos propuestos en el Algoritmo-Rayuela, aunque alterando su orden o adaptando la secuencia según fuera conveniente, siempre habilitando el intercambio de ideas y las discusiones, favoreciendo la exposición de diferentes estrategias de abordaje o de caminos alternativos.

Los testimonios recabados luego del primer examen parcial indican que la estrategia fue valorada positivamente por los estudiantes, quienes remarcaron su utilidad para ordenar la información, para orientar sobre los pasos a seguir y para favorecer la comprensión por parte de un tercero de lo producido. Finalmente, se ha notado que los alumnos logran desarrollar las resoluciones de manera ordenada con sus correspondientes gráficos, posponen el reemplazo por valores numéricos hacia la fase final y presentan sus trabajos de manera clara y completa, mostrando la adquisición paulatina de capacidades comunicacionales.

## Referencias